# Small scale CMB fluctuations
# as a probe of the mass of Dark Matter particles

Andrei G. Doroshkevich [1,2] and Raffaella Schneider [2,3]

[1]Keldysh Institute of Applied Mathematics, Russian Academy of Sciences, 125047 Moscow, Russia

[2]Theoretical Astrophysics Center, Blegdamsvej 17, 2100 Copenhagen Ø, Denmark

[3]Dipartimento di Fisica, Universita' di Roma " La Sapienza", Piaz.le Aldo Moro 2, 00185 Rome, Italy

# ABSTRACT

The CMB anisotropy on arc second range is examined to test the power spectrum of perturbations in the small scale region and, in particular, to estimate the mass of dominant dark matter particles. It is shown that for the simplest evolutionary history with standard recombination, three and four beam observations could discriminate the mass of dark matter particles in the interval $0.5 KeV \leq M_{DM} \leq 4 KeV$ with an antenna beam (0.5 - 0.25) arc minute with amplitude $\approx 10^{-7}$. This interval is the most interesting one for the problem of Large Scale Structure formation and evolution.

*Subject headings:* cosmology: theory - cosmic microwave background - large scale structure of the Universe - dark matter

## 1. Introduction

The investigation of the power spectrum of primordial perturbations is now a key problem of observational cosmology. The recent detection of large scale cosmic microwave background (CMB) anisotropies by COBE (Smoot et al. (1992)) has provided us with a unique opportunity to probe the properties of primordial perturbations in large and intermediate scales. These data allow to estimate the amplitude and slope of power spectra in the large scale area (Gorski et al. (1994), Bond (1994)). At the same time, the complex analysis of motion and spatial distribution of galaxies has provided valuable information about power spectrum at smaller scales (Peacock & Dodds (1994), da Costa et al. (1994)). These results show the large progress that has been made in observational cosmology during the last years.

It can be expected that in the next years a strong effort will be concentrated around CMB anisotropies on scales from several arc-minutes to few degrees. Actually, these investigations could provide valuable information about the most fundamental cosmological parameters, namely the initial power spectrum, the Hubble constant, the cosmological constant, the baryon density, the ionisation history, the dynamic of recombination, etc.

However, the huge potential of CMB investigations is not still exhausted with the problems mentioned above. In this paper we would like to proceed further and to reveal the close connection of small-scale CMB anisotropy with a fundamental open issue of modern cosmology that is the nature, mass and properties of the dominant dark matter particles or, in other words, the shape of the power spectrum on the smallest scales. This area of the spectrum is important in itself for a reconstruction of the evolution history of the universe, for a description of processes of galaxy formation and as an important test of inflation. And, what is perhaps especially interesting, the shape of the power spectrum in the short wave area is very sensitive to the mass of the dominant type of dark matter (DM) particles,

$M_{DM}$.

Particle physics gives us a set of possible candidates for DM particles. First of all there are three types of neutrinos, the expected masses of which are smaller than 10 eV. Then there is the massive long-lived majoron with mass in the KeV range (Akhmedov et al. (1993), Berezinsky & Valle (1993)). Finally, some fraction of DM density can be related to the neutralino with a mass which is more than 20 GeV. It is a very important task to discriminate among these versions and to single out the mass of the dominant type of DM particles.

There are few possible ways to detect DM particles. The most direct method is, of course, the laboratory test of some of their possible properties (for example, with decays of particles). However, in this way we can only confirm (or reject) some expectations of the theory which indeed could be not so certain. A valid, though not simpler, alternative is the investigation of the power spectrum in the small scale region. This method provides a certain information about the mass of DM particles which is, of course, the most important parameter. The analysis of galaxies distribution cannot provide us with the necessary information because small scale perturbations disappeared during the nonlinear evolution of Large and Super Large Scale Structure (LSS and SLSS). However, some information about the small scale area of the power spectrum is retained in small-scale CMB anisotropy and can be, in principle, set off with three and four beam observations at angular scales of about 0.5 - 0.25 arc-minutes.

The current more popular DM models are Cold Dark Matter (CDM, Blumenthal et al. (1984)) and Cold + Hot Dark Matter (CHDM, Klypin et al. (1993)). However, recent theoretical investigations of LSS formation and evolution (Doroshkevich (1995), Doroshkevich et al. (1995) hereafter DFGMM) show that the best description of these processes is provided by the Warm Dark Matter (WDM) model with the transfer function proposed by Bardeen et al. (1986) (hereafter BBKS). It describes the continual transformation from Hot Dark Matter to CDM and allows to obtain typical scales of LSS and SLSS which are in best agreement with the observational estimates (Buryak et al. (1994), Doroshkevich & Tucker (1995)). The WDM model provides that the structures form during a reasonable redshift interval $z \approx 5 - 20$ for particles masses in the range $M_{DM} \approx 0.5 - 4 KeV$. Therefore the mass of DM particles predicted by the theory of structure formation is close to the possible mass of majoron, $m_j$, evaluated by Berezinsky & Valle (1993). However, the upper limit found in this paper, $m_j \leq 1.34 h^2 KeV$ ($h = H_0/100$ km/s/Mpc is a dimensionless Hubble constant), is smaller than the values expected from dynamical models. Although this problem needs to be investigated in more detail, a comprehensive analysis of some consequences of the WDM model becomes important.

The differences between WDM and CDM models are significant only in the small scale area. On large and intermediate scales all the predictions of both models are the same. Therefore, the analysis of small-scale CMB anisotropy is a unique way to test the WDM model and to provide observational estimates of the mass of DM particles.

In this paper we examine a very simple model which assumes a standard recombination at a redshift $z_{rec} = 1000$ without secondary ionisation up to small redshifts. Our analysis is based on an approximate semi analytical description of the radiation transfer function because here our main aim is to test roughly the proposed approach. The numerical estimates listed below show that this way seems to be promising enough and deserves to be examined thoroughly. Further steps of the analysis can include on one hand a more comprehensive investigation of the process of LSS and SLSS formation which will allow to find a more profitable range of parameters of DM particles. On the other hand, detailed numerical simulations of CMB anisotropies could be performed for different cosmological scenarios.

The paper is organized as follows. In Sec. 2 we provide a short description of the theory of LSS and SLSS formation and evolution with WDM model in order to emphasize some of the principle differences between WDM and the scale-free CDM model. This description can be considered as a ground to prefer the first one as the most realistic DM model. In Sec. 3 small-scale CMB anisotropy for WDM and CDM models is investigated. In Sec. 4 we discuss and summarize our conclusions.

## 2. LSS and SLSS formation and evolution with a WDM model

A quantitative, albeit approximate, description of the formation and evolution of LSS and SLSS has been recently proposed by Buryak et al. (1994), Doroshkevich (1995) and DFGMM. This theory starts from the Zel'dovich approximation (Zeldovich (1970)), the great potential of which has been repeatedly demonstrated, and shows that the evolution of LSS can be described by a few approximate expressions involving simply moments of the power spectrum. In particular, it establishes the relationship of characteristic measures of LSS and SLSS and of their evolutionary behavior with the power spectrum. Some aspects of the formation of SLSS have already been discussed by Demiański & Doroshkevich (1992) and Buryak, Demiański & Doroshkevich (1992) though the evolution of LSS could be more useful in discriminating the many DM models under study today. For instance, it allows to reject the first and simplest HDM model.

Although there is little presently available information about LSS evolution, the acquisition of more and better quasar absorption spectra and their detailed and careful analysis provides an exciting possibility for the observation of the evolution with time of large-scale structure in the Universe up to redshifts $z \approx 5$.

The problem of SLSS and LSS formation and evolution can be formulated in terms of typical scales or, in other words, in terms of moments of the power spectrum. BBKS considered the set of even spectral moments, namely

$$\sigma_j^2 = \int \frac{k^2 dk}{2\pi^2} p(k) k^{2j} \qquad (2.1)$$

and used mainly three moments, $\sigma_0$ which is identical to the amplitude of perturbations and the values

$$<k^2> \equiv \sigma_1^2/\sigma_0^2 \qquad <k^4> \equiv \sigma_2^2/\sigma_0^2 \qquad (2.2)$$

which play an important role in the process of structure formation.

The expressions for $\sigma_0$ and $<k^2>$ are divergent for a *scale-invariant* CDM spectrum. In reality, there is a cut-off at large $k$ and according to WDM models (BBKS) this cut-off is directly related to the mass of the DM particles dominating the universe, i.e. $k_{max} \propto M_{DM}(M_{DM}/\Omega_m h^2)^{1/3}$. For neutrino types of WDM models a similar mass dependence for the damping scale, $k_{max} \propto M_{DM}$, can be found (BBKS).

In this paper we will concentrate our attention around three DM models with transfer functions

$$T_{CDM} = \frac{ln(1 + 2.34q)}{2.34q} \left( 1 + 3.89q + (16.1q)^2 + (5.46q)^3 + (6.71q)^4 \right)^{-1/4} \qquad (2.3a)$$

$$T_{WDM}^{(1)} = (1 + 1.7q + (4.3q)^{3/2} + q^2)^{-1} \times \exp(-0.5qR_f - 0.5(qR_f)^2) \qquad (2.3b)$$

$$T_{WDM}^{(2)} = (1 + 1.7q + (4.3q)^{3/2} + q^2)^{-1} \times \exp(-0.5qR_m - 0.5(qR_m)^2) \qquad (2.3c)$$

$$q = k/(\Omega_m h^2 Mpc^{-1}) \quad R_f = 0.2(\frac{\Omega_m h^2}{M_{DM}})^{4/3} \quad R_m = 0.26\frac{\Omega_m h^2}{M_{DM}}$$

where $M_{DM}$ is in KeV. The first model is simply the scale free CDM model, the second one is BBKS WDM model and the third one is a neutrino type WDM model. The two WDM models differ only for the mass dependence of the damping scale, namely $R_f$ and $R_m$. The numerical coefficients for the damping scales depend weakly on the particles composition and need to be corrected for various cosmological models.

Strictly speaking, the following theoretical analysis, based on the Zeldovich' theory, can be performed only for the WDM models and not for the scale free CDM model. This remark becomes clearly superfluous when referred to numerical simulations since in this case an artificial cut-off for the power spectrum is introduced. It is interesting that both theoretical predictions and simulated results for the DM spatial distribution are insensitive to the substitution of the CDM model (with a short wave cut-off) to WDM models. However, this is no longer valid for the spatial distribution of baryonic matter and galaxies (see, e.g., Fong et al. (1995)).

Following Buryak et al. (1994) and DFGMM we will consider the distribution of LSS elements along a random straight line (core-sampling approach). In this manner the structure elements can be strictly defined both in theory (Zeldovich' pancakes) and in simulated and observational catalogues. The one dimensional spatial distribution of structure elements is found to be Poissonian. Therefore, the mean separation between the elements is the main physical characteristic of the structure and provides the quantitative description of structure evolution. Analysis performed by Buryak et al. (1994) and DFGMM shows that the mean 1D separation of LSS elements rapidly decreases during the first period of the active structure creation. Later, the sequential merging of LSS elements leads to a slow increase of the mean separation. Here, we present without proof the main quantitative relations of the theory, referring the readers to the original papers for more details.

According to Zeldovich' theory, the typical separation of structure elements along a random straight line, $l_{cr}$, during the active phase of structure formation is defined through two first moments of the power spectrum as follows:

$$l_{cr}^{-1} \simeq r_c^{-1} \left[ \frac{8}{3\pi^2} \sqrt{\frac{15}{7}} \left( 1 + y^2/2 \right) \exp\left(-y^2/2\right) \right] \leq 0.4 \, r_c^{-1}, \qquad (2.4)$$

$$y = (1+z)/(1+z_{cr}) = (1+z)\sqrt{5}/\sigma_0, \quad r_c^2 = 3./ < k^2 >$$

where $z$ is the redshift and $z_{cr}$ is the 'characteristic epoch' for the creation of structure elements (Doroshkevich (1984), Doroshkevich (1989)). The values $z_{cr}$ and $r_c$ are presented in Table I for CDM and WDM models.

A more detailed theoretical analysis (DFGMM) shows that other even moments are equally important for structure evolution. Namely, for CDM like broad band power spectra with Harrison-Zeldovich asymptotic

$$p(k)/2\pi^2 = (9/16)l_{max}^4 kT^2(k) \tag{2.5}$$

$$l_{max}^2 = \sqrt{\frac{16}{15}Q_2 R_h^2}, \quad l_{max} \approx 15.2h^{-1}Mpc$$

with a quadrupole anisotropy $Q_2 = (17/2.73) \times 10^{-6}$ and a horizon $R_h = 6000h^{-1}Mpc$, the LSS evolution due to random motion and merging can be approximately described by the following equation:

$$l_{dis}^2 \approx \frac{2}{(1+z)^2} \frac{1}{2\pi^2} \int_0^\infty p(k)G(kl_{dis})dk \tag{2.6}$$

$$G(x) = x^{-3} \int_0^x u^2 \exp(-u^2)du$$

where $l_{dis}$ again is a characteristic scale of the separation of LSS elements along a random straight line. A simpler approximation to this equation is

$$l_{dis}^2 \approx \frac{3}{4} \frac{l_{max}^4(1+z)^{-2}}{\sqrt{l_0^4 + l_{max}^4(1+z)^{-2} + l_0^2}}, \tag{2.7}$$

$$l_0^{-2} = \int_0^\infty kT^2(k)dk$$

Thus, the structure evolution can be described through the typical scale $l_0$ which is, of course, the -1 moment of the spectrum,

$$\sigma_{-1} = (9/16)l_{max}^4 l_0^{-2} \tag{2.7}$$

Values of $l_0$ for CDM and WDM models are listed in Table I. It is important that during the second evolutionary period (merging), the strong differences between the predictions of CDM and WDM models are eliminated and the main quantitative results for the present matter distribution are the same (see $l_{LSS}(z = 0)$, $l_0$ and $L_0$ in Table I).

For CDM and WDM[1], the LSS evolution according to relations (2.3) and (2.7) is shown in Fig.1 where the values

$$l_{LSS} = max(l_{cr}, l_{dis}) \qquad (2.8)$$

are presented vs. redshift $z$. The typical redshifts $z_{min}$ when $l_{cr} = l_{dis}$ are listed in Table I.

A similar description can be given for the evolution of SLSS. In this case, the typical scale $L_0$ is defined through the spectral moments $\sigma_{-1}$ and $\sigma_{-2}$ (Buryak et al. (1994), DFGMM) and its values are presented in Table I for CDM and for the two versions of the WDM model.

Numerous theoretical investigations and numerical simulations demonstrate clearly that the LSS is an *intermediate* asymptotic of the general process of formation and evolution of cosmic structures. Generally, structures are formed through rapid collapse of matter along one axis to form a Zel'dovich 'pancake' and with less rapid collapse along a second orthogonal axis to form a 'filament', that is, creating a 'ridge' of higher over-density within the pancake. They then evolve not only by merging together according to the (2.6), but also through a general motion of matter along the filaments towards the branch points of this 'general network structure'. This latter process destroys the regular LSS elements, creating the sites for the formation of the richest clumps of matter. We refer the reader to Shandarin & Zeldovich (1989) for an excellent review of much of this work. Later these processes were discussed in more detail by Kofman et al. (1992) and by Buryak et al. (1994).

Quantitative estimates of the structure evolution were tested recently with six numerical simulations performed for CDM and Broken Scale Invariance power spectra with various computational boxes (Doroshkevich et al. (1995)). These simulations cover an interval $1 \leq (1 + z_{cr}) \leq 10$ and, thus, allow to test various periods of structure evolution. This analysis shows that developing LSS can be found in the redshift interval

$$0.25 \leq \frac{1 + z}{1 + z_{cr}} \leq 2.5 \qquad (2.9)$$

At present, the farthest galaxies are found up to redshifts $z \simeq 3 - 4$ (see, e.g., Giavalisco et al. (1995)) and regular LSS elements are observed in deep surveys (Buryak et al. (1994), Doroshkevich & Tucker (1995)). At the same time, the quiet evolution of LSS described by equation (2.6) is observed through the analysis of quasar absorption spectra (Doroshkevich & Turchaninov (1995)). These data allow to conclude that the quiet evolution of LSS took place at least for $0 \leq z \leq 5$ and from equation (2.9) we can infer the corresponding values of $z_{cr}$. These are found to lie within the interval $1 + z_{cr} \approx 2 \div 4 - 5$ and correspond to $WDM^{(1)}$ models with $M_{DM} \approx 0.5 - 2 KeV$ for the transfer function (2.3b) and to $WDM^{(2)}$ models with $M_{DM} \approx 0.5 - 3 KeV$ for the transfer function (2.3c).

A more detailed analysis shows some dispersion in the moments of structure creation that gives rise to some uncertainties in the estimates of $z_{cr}$ from (2.9) and $M_{DM}$. For this reason, we investigate a wider range of values for $M_{DM}$, namely $0.1$KeV$\leq M_{DM} \leq 8$KeV. Additional theoretical investigations and corresponding numerical simulations need to be performed in order to narrow the intervals for $z_{cr}$ and $M_{DM}$.

A similar approach, based on the analysis of dynamical processes of LSS evolution, is an important part of theoretical cosmology allowing to employ the observational data for the spatial galaxy distribution together with numerical simulations. Through this method some important constraints can be set on possible parameters of DM particles and on the power spectrum. However, this is still an indirect way because any information about the actual processes of structure formation cannot be achieved even with a modern astronomical technique. At the present time, the farthest quasars can be found with a redshift $z \approx 4.5 - 5$ only and even such information is a very poor one. Thus, to investigate these processes we need to have some alternative ways which could provide a direct information about the small scale range of the power spectrum and, therefore, about the early processes of galaxy and structure formation.

Such way is the investigation of CMB anisotropies on sub-minute scales.

## 3. Small-scale CMB anisotropy

The theoretical analysis of CMB anisotropies was fulfilled by many authors and now the general behavior of anisotropies is well known for most models. Detailed numerical computations have already been discussed in many papers (e.g. Peebles & Yu (1970), Wilson & Silk (1981), Bond & Efstathiou (1984), Bond & Efstathiou (1987), Vittorio & Silk (1984),White, Scott & Silk (1994)) and provide the best comparison of a specific model with observational data. However, as a first step towards new investigations of a model, an analytical (or semi analytical) method is sufficiently accurate and to be preferred. This kind of approach requires the use of an analytical fit for the radiation transfer function $\Delta_{rad}(k, \mu)$ which relates the primordial power spectrum of matter perturbations to the CMB anisotropies.

The simplest analytical expression for the radiation transfer function was proposed by Doroshkevich (1988) and Atrio-Barandela & Doroshkevich (1994). Here a modified version of the transfer function was employed. It takes into account the variation of metric parameters during the period of equality of relativistic and non relativistic matter density, as suggested by Hu & Sugiyama (1994), and provides a valid description of small and large-scale perturbations together with the first Doppler peak but decreases the amplitude of second and further peaks. Since our analysis is concentrated around small-scale anisotropy, this improved form of the transfer function is sufficiently accurate for our aims. A more accurate but significantly more cumbersome semi analytical approach has been proposed by

Hu & Sugiyama (1994).

At present, the most popular method to describe CMB anisotropies is through the squares of the coefficients of spherical harmonics (the $C_l$'s), which is very convenient for large and intermediate angular scales (see, e.g., Bond (1994)). However, for small-scale anisotropy the description with an angular correlation function is more suitable (for the subdegrees scale such approach was discussed in detail by Jorgensen et al. (1995)). It is interesting that it can also be performed in terms of spectral moments or typical scales. The specific damping of CMB anisotropies during the process of recombination requires to include in the analysis the odd moments as well. Thus, the coherence angle, $\theta_c$, can be related to the $\sigma_{-3/2}$ spectral moment and, what is more interesting for us, the expressions for a three and four beam experimental set up relate to $\sigma_{-1/2}$ and $\sigma_{1/2}$ respectively.

## 3.1. The approximate description of small-scale CMB anisotropy

As it is known (e.g. Bond & Efstathiou (1987), Efstathiou (1988)), the angular correlation function for the Harrison - Zeldovich primordial spectrum can be written as follows:

$$C(\theta, \theta_s) = \frac{12}{5} \left(\frac{Q_{rms}}{T_0}\right)^2 \int_0^\infty \frac{dk}{k} T^2(k) \int_0^1 d\mu \Delta_{rad}^2 e^{-S_d - S_s} J_0(y) \qquad (3.1)$$

$$y = 2kR_h sin(\theta/2)(1 - \mu^2)^{1/2}$$

$$S_d = (kd_r\mu)^2 \quad S_s = (kR_h\theta_s)^2(1 - \mu^2)$$

where $\theta$ is the switch angle between the beams, $T(k)$ is the usual matter transfer function, the radiation transfer function $\Delta_{rad}^2(k, \mu)$ describes the transformation of the power spectrum to the radiation field, $J_0(x)$ is the Bessel function, the value $d_r \approx 0.1R_{rec}$ gives the damping of anisotropies during the recombination, $R_h$ and $R_{rec}$ are the present horizon and the horizon at the moment of recombination and we approximate the finite antenna beam size as a Gaussian of angular width $\theta_s$ .

The function $\Delta_{rad}^2(k,\mu)$ can be written as

$$\Delta_{rad}^2(k,\mu) = A_0^2(k) + 3\mu^2 B_0^2(k)$$

$$A_0 = -1 + \frac{u^2}{6a_r + u^2}(1 - 4cos(u - u_{eq}) + 3cosu + 3a_r) \qquad (3.2a)$$

$$B_0 = \frac{u^2}{6a_r + u^2}(6a_r/u + 4sin(u - u_{eq}) - 3sinu) \qquad (3.2b)$$

with $u = kR_J|_{z=z_{rec}}$, $a_r = (4\rho_b/3\rho_\gamma)(1 + z_{rec})^{-1} = 31h^2\Omega_b$, $u_{eq} = kR_J|_{z=z_{eq}}$, $z_{eq} = \rho_m/\rho_\gamma = 40\Omega h^2 z_{rec}$ where $\Omega_b$ is the dimensionless baryon density, $z$ is redshift and $z_{rec} = 1000$ corresponds to the standard recombination. Clearly $\Delta_{rad}^2 \approx 1$ for small $k$ and $\Delta_{rad}^2 \approx 9a_r^2 + 12.5(1 + 3\mu^2)$ for large $k$. The expression (3.2) matches the relation proposed by Atrio-Barandela & Doroshkevich (1994) for $u_{eq} = 0$. Using hypergeometric functions we obtain a better agreement with numerical computations in the area of the first Doppler peak. For larger $u$ the trigonometric approximation is sufficiently accurate.

The function $\Delta_{rad}^2(k,\mu)$ (3.2) does not include the diffusion damping of radiation which can be essential for small scale perturbations because the typical scale of such damping is $d_D \approx 1.5(\Omega_b h)^{-1/2}h^{-1}$ Mpc. Let us note, however, that this factor can affect only some quantitative estimates but can not cancel the considered effect because here we have to deal with *forced* perturbations as the perturbations in the DM component generate perturbations in the baryons and CBR. Therefore, the diffusion of photons leads to a strong depression of the oscillatory terms in (3.2) and decreases the large $k$ asymptotic of the function $\Delta_{rad}$ approximately up to $9a_r^2$ and a strong dependence of the final results on $\Omega_b$ appears. This effect is more important for small values of $\Omega_b$ when the photon mean free-path increases strongly and the hydrodynamical treatment becomes incorrect in the smallest scales. In this case the kinetic analysis must be used and this effect needs to be considered further.

The amplitude of the correlation function, $C(0,\theta_s)$ can be written as follows:

$$C(0,\theta_s) = \frac{12}{5}\left(\frac{Q_{rms}}{T_0}\right)^2 C_0$$

$$= \frac{12}{5} \left( \frac{Q_{rms}}{T_0} \right)^2 \int_0^\infty \frac{dk}{k} T^2(k) \int_0^1 d\mu \Delta_{rad}^2 exp(-S) - monopole - dipole \qquad (3.3)$$

with $S = S_d + S_s$.

For small switch angles, $\theta \ll \theta_c$, the expression (3.1) can be rewritten in the following standard form:

$$\frac{C(\theta, \theta_s)}{C(0, \theta_s)} = 1 - \frac{\theta^2}{2!\theta_c^2} + \gamma_1 \frac{\theta^4}{4!\theta_c^4} - \gamma_2 \frac{\theta^6}{6!\theta_c^6} \qquad (3.4)$$

where $\theta_c$ is the coherence angle and $\gamma_1$ and $\gamma_2$ are dimensionless factors describing the short-wave part of the power spectrum (see, e.g., Bond & Efstathiou (1987)). If damping effects are not taken into account and $\theta_s = 0, d_r = 0$ in (3.1) then $\gamma_1$ is proportional to the zero spectral moment, $\sigma_0$, and is divergent for a CDM power spectrum. With the damping during recombination, $\gamma_1$ retains a finite value even for CDM power spectra and it is proportional to the moment $\sigma_{-1/2}$. A further expansion of the correlation function in the series for a standard CDM power spectrum can be performed only with a finite beam size (Bond & Efstathiou (1987)) since for $\theta_s \to 0$, $\gamma_2 \to \infty$.

For WDM power spectra the exponential cut-off of the transfer function $T(k)$ changes the factors $\gamma_1$ and $\gamma_2$ in (3.4) and the effect depends on the mass of DM particles, $M_{DM}$. Therefore, this way can be used for the determination of the shape of the power spectrum in the short-wave area and, in principle, for the determination of $M_{DM}$. For WDM spectra the factor $\gamma_2$ is convergent even for a point-like antenna. However, even in this case the beam smearing can have very important consequences.

With two beam observations

$$D_2 = \frac{< (T_1 - T_2)^2 >}{T_0^2} = 2 \left( C(0, \theta_s) - C(\theta, \theta_s) \right) = C(0, \theta_s) \frac{\theta^2}{\theta_c^2} \qquad (3.5a)$$

it is possible to find only the coherence angle $\theta_c$ and, thus, other experimental arrangements are more suitable for our aims. The three beam observations (Uson & Wilkinson (1984))

$$D_3 = \frac{< (T_1 + T_3 - 2T_2)^2 >}{4T_0^2} = 1.5C(0, \theta_s) - 2C(\theta, \theta_s) + 0.5C(2\theta, \theta_s)$$

$$= C(0, \theta_s) \frac{\gamma_1}{4} \frac{\theta^4}{\theta_c^4} \qquad (3.5b)$$

are more sensitive to the small-scale region of the power spectrum and they allow to determine the factor $\gamma_1$. And, of course, the four beam observations

$$D_4 = \frac{< (T_1 + 3T_3 - 3T_2 - T_4)^2 >}{4T_0^2} = 5C(0, \theta_s) - 7.5C(\theta, \theta_s) + 3C(2\theta, \theta_s) - 0.5C(3\theta, \theta_s)$$

$$= C(0, \theta_s) \frac{\gamma_2}{4} \frac{\theta^6}{\theta_c^6} \qquad (3.5c)$$

are the most effective way to test the possible mass of DM particles through the factor $\gamma_2$.

## 3.2. The two and three beam observations

For the further analysis it is convenient to introduce three characteristic lengths related to the transfer function as follows:

$$l_{(-3)}^{-1} = \int_0^\infty dk T^2 \; e^{-(kd_s)^2} \qquad (3.6a)$$

$$l_{(-1)}^{-3} = \int_0^\infty dk k^2 T^2 \; e^{-(kd_s)^2} \qquad (3.6b)$$

$$l_1^{-5} = \int_0^\infty dk k^4 T^2 \; e^{-(kd_s)^2} \qquad (3.6c)$$

where $d_s = R_h \theta_s$ and we include in the lengths definition the smoothing effect of the antenna. For small values of $\theta_s$ (and $d_s$) the influence of this factor on $l_{(-3)}$ and $l_{(-1)}$ is negligible and these lengths are well defined even for a point-like antenna. On the contrary the influence of the antenna is extremely important for the third length, $l_1$, which is divergent for the scale-free CDM power spectrum and is very sensitive to the mass of DM particles, $M_{DM}$, for the WDM spectrum.

The coherence angle can be written as ($d_s < d_r$ for $\theta_s < 10'$)

$$\theta_c^2 = \frac{4}{\sqrt{\pi}} \frac{\sqrt{d_r^2 - d_s^2}}{R_h^2} C_0 \frac{l_{(-3)}}{(90\Omega_b h^2)^2 + 25/2} \qquad (3.7)$$

The amplitude of the correlation function $C_0$ and the characteristic length $l_{(-3)}$ are sensitive mainly to the comparatively large-scale range of the power spectrum where possible distortions of its primordial shape are not significant. Moreover, this interval of wave lengths can be described and investigated with more accuracy through the $C_l$'s (see, e.g., Bond (1994)). This relation demonstrates the remarkable stability of the coherence angle relatively to various models because the Doppler peak range has a moderate influence on $C_0$ and $l_{(-3)}$. Therefore we can conclude that the coherence angle $\theta_c$ is mainly a function of the Hubble constant (as $l_{(-3)} \propto h^{-2}$) and the damping factors so that two beam experiments promise to give important information about the damping during recombination and about possible processes of secondary ionisation.

A three beam configuration seems to be more interesting for us as it is more sensitive to the small-scale range of power spectrum. Actually,

$$\gamma_1 = \frac{3\sqrt{\pi}}{16} \frac{R_h^4}{\sqrt{d_r^2 - d_s^2}} \frac{l_{(-1)}^{-3}}{(90\Omega_b h^2)^2 + 25/2} \frac{\theta_c^4}{C_0} \tag{3.8}$$

and with known coherence angle it allows to estimate the typical scale $l_{(-1)}$.

The results of calculations of $\gamma_1$ for two WDM models together with $\theta_c$ are presented in Fig.2 vs. the mass of DM particles, $M_{DM}$ for $h = 0.5$ and $\Omega_b = 0.1$. As it was expected, the coherence angle $\theta_c$ is a very weak function of $M_{DM}$. At the same time, $\gamma_1$ increases of a factor three in the range of masses $0.1 - 8\ KeV$ and this effect, in principle, can be set off with three beam observations.

It is necessary to test the influence of the antenna damping as well. Our estimates of the antenna damping are also presented in Fig.2 for three possible beam sizes, $\theta_s = 0.5', 0.25'\ \&\ 0.1'$. They show that with an antenna angle of $\theta_s = 0.5'$ we are able to test only the upper limit of $M_{DM} \leq 0.5\ KeV$ and more narrow beams need to be employed to investigate the most interesting range of $0.5\ KeV \leq M_{DM} \leq 2 - 4\ KeV$. The antenna angle $\theta_s = 0.25'$ seems to be sensitive enough up to $M_{DM} \leq 2\ KeV$ whereas $\theta_s = 0.1'$ covers the full interval of expected masses, $M_{DM} \leq 8 \ KeV$.

### 3.3. The four beam observations

The four beam experiment had been proposed by Jorgensen et al. (1995) to investigate the sub degrees range of spectrum (the Doppler peak area). Here we propose to use such approach with a different geometrical configuration.

Clearly a four beam experiment is further more sensitive to the small-scale range of the power spectrum and is further more suitable for the test of $M_{DM}$. As it was noted above, for high order terms in the series (3.4) the coefficients become strong functions of the beam size. However, for the same beam size the four beam experiment is more sensitive to $M_{DM}$ as the factor

$$\gamma_2 = \frac{5\sqrt{\pi}}{32} \frac{R_h^6}{\sqrt{d_r^2 - d_s^2}} \frac{l_{(1)}^{-5}}{(90\Omega_b h^2)^2 + 25/2} \frac{\theta_c^6}{C_0} \tag{3.9}$$

include the characteristic length $l_1$.

In Fig.3a and 3b we plot $\gamma_2$ as a function of the switch angle $\theta$ for two beam sizes, $\theta_s = 0.25'$ & $0.5'$ and for $h = 0.5$ and $\Omega_b = 0.1$. We investigate two WDM models for different values of $M_{DM}$ within the interval $0.1 \ KeV \leq M_{DM} \leq 8 \ KeV$ and the scale-free CDM model.

These Figures demonstrate clearly the great potential of four beam observations for the study of small-scale regions of power spectra and, thus, for the detection of the mass of DM particles, $M_{DM}$. As it could be expected, for small masses, $M_{DM} \leq 0.5 \ KeV$, the function $\gamma_2$ is approximately *const.* what is a natural result of the strong damping of power spectra in the short-wave region. However, as $M_{DM}$ grows the area of application of expressions (3.4) & (3.9) reduces. When $M_{DM}$ becomes $\geq 1 \ KeV$, the results of the computations for

WDM models match the CDM case on a common asymptotic of the form,

$$\gamma_2 \approx \kappa(\theta_c/\theta)^3 \qquad (3.10)$$

with $1 \leq \kappa \leq 2$. In this case expression (3.4) is no longer valid and $\gamma_2$ depends strongly on $\theta$. Nevertheless, the differences between the predictions of various models are large enough to be discriminated with narrow antenna beams.

Fig.3a and 3b demonstrate that the influence of the antenna beam is not very significant for $\theta_s \approx 0.5', 0.25'$ and the upper limit of the mass range that can be investigated depends more on the model itself. For WDM[1] the limit is set by $M_{DM} \leq 2 \ KeV$ while for WDM[2] masses $M_{DM} \approx 4 \ KeV$ can be discriminated. Finally, it is possible to show that for both models the interval of masses $M_{DM} \leq 1 \ KeV$ can be investigated even with a beam size $\theta_s \approx 1'$.

These parameters clearly exceed the available ones. However, this is the most direct way to test the small-scale range of the power spectrum and, thus, to solve many key problems of modern cosmology.

## 4. Discussion

In this paper we attempt to demonstrate that the huge potential of CMB anisotropies is not exhausted with the most popular problems as the initial power spectrum, the Hubble constant, the cosmological constant, the baryon density, the ionisation history, the dynamic of recombination, etc. At present, sub-minute angular scales are not as popular as the sub-degree ones. However, our analysis shows the significant potential of investigating fluctuations at the smallest angular scales with three and especially four beam experiments.

In the previous Sections we have stressed the close connection of small-scale CMB anisotropy with the nature and properties of dominant DM particles, and therefore with the fundamental problem of the formation of various cosmic structures like galaxies, clusters of galaxies and elements of Large and Super Large Scale Structure. This way seems to be the most realistic to reveal many crucial problems of modern cosmology and modern particle physics because the nature of DM particles is one of the fundamental issues in which the interests of both branches of science are intersected.

Our analysis has a rather preferential character. The main aim of this paper is to present under the simplest assumptions the principle abilities of the proposed approach rather than to give an exhaustive discussion of all the details. Thus, many important problems were only mentioned and undoubtedly require further investigations. The first of all is the effect of photon diffusion, which will change the quantitative estimates and can introduce a strong dependence of the results on the baryon density, $\Omega_b$. Special attention should also be devoted to the possibility of a non-standard recombination since secondary ionisation will strongly increase the damping scale. In these cases, the hydrodynamical approach used above is no longer valid (especially for small values of $\Omega_b$) and the kinetic analysis needs to be performed. However, for intermediate values of $\Omega_b$ ($\Omega_b \geq 0.03$) a more accurate analysis can only rectify but certainly not lose the strong dependence of the results of four beam observations on the mass of dark matter particles, $M_{DM}$.

Random noise is especially important for narrow beam observations and was discussed in detail during the eighties. Since we do not touch this very specific and complicated topic, we refer to the excellent reviews of Partridge (1988), Readhead et al. (1992) and White, Scott & Silk (1994).

The problem of DM particles is very important in itself and it is sufficient ground to develop further the analysis of CMB anisotropy on sub-minute scales. However, here we would also like to emphasize the strong connection of this problem with many key issues of modern cosmology, like the fundamental process of galaxy formation and many enigmatic results of quasar observations. Moreover, it relates to the old problem of the evolutionary history and spatial distribution of various components of matter, namely luminous, baryonic and dark matter. Any progress in these directions requires the collection of more and more detailed information about the small-scale range of the power spectrum and, thus, about the mass and other properties of DM particles.

In Section 2 some arguments were given to show this close connection through both qualitative and methodological considerations. In particular, it is significant that fundamental properties of LSS and SLSS as well as the main parameters of small-scale CMB anisotropy can be described in a similar way using moments or characteristic lengths of the power spectrum. Thus, the initial perturbations of gravitational potential provide large-scale CMB anisotropy and the galaxies concentration in SLSS elements. On the other hand, small-scale CMB anisotropy and the process of galaxy formation are described by the short-wave area of the power spectrum. This approach emphasizes the common physical ground and complementary character of various observational phenomena in different branches of cosmology. At the same time it shows that only a complete analysis of CMB anisotropy in a wide range of angular scales could provide a reliable solution of all the above mentioned problems.

This paper was supported in part by Danmarks Grundforskningsfond through its support for an establishment of Theoretical Astrophysics Center. A.D. wishes to acknowledge support from the Center of Cosmo-Particle Physics, Moscow.

| | $M_{DM}$ | $L_0$ | $l_{LSS}(z=0)$ | $l_0$ | $r_c$ | $1+z_{cr}$ | $1+z_{min}$ |
|---|---|---|---|---|---|---|---|
| CDM | $\infty$ | 31 | 3.4 | 7.3 | 0 | $\infty$ | $\infty$ |
| WDM[1] | 0.1 | 40.0 | 4.60 | 7.8 | 1.7 | 1.2 | 0.9 |
| | 0.5 | 28.5 | 3.75 | 6.2 | 0.26 | 4.0 | 5.6 |
| | 1.0 | 27.9 | 3.65 | 6.1 | 0.11 | 6.2 | 10 |
| | 2.0 | 27.4 | 3.65 | 6.0 | 0.05 | 8.9 | 16.7 |
| | 4.0 | 27.2 | 3.60 | 6.0 | 0.03 | 12.3 | 26.3 |
| | 8.0 | 27.1 | 3.60 | 5.9 | 0.02 | 16.1 | 39.2 |
| WDM[2] | 0.1 | 39.3 | 4.30 | 7.6 | 1.6 | 1. | 0.8 |
| | 0.5 | 30.1 | 3.65 | 6.3 | 0.39 | 3.2 | 3.8 |
| | 1.0 | 28.5 | 3.62 | 6.1 | 0.21 | 4.5 | 7.1 |
| | 2.0 | 27.5 | 3.60 | 6.1 | 0.11 | 6.2 | 10.1 |
| | 4.0 | 27.4 | 3.60 | 6.0 | 0.06 | 8.2 | 15.3 |
| | 8.0 | 27.3 | 3.60 | 6.0 | 0.03 | 10.6 | 22.5 |

Table 1: The $M_{DM}$ is in KeV and all the distances are presented in $(\Omega h^2)^{-1}$ Mpc.

# REFERENCES

Akhmedov, E.Kh., Berezhiany, Z.G., Mohapatra, R.N. & Senjanovic, G., 1993, Phys.Lett. **B 299** 90

Atrio-Barandela, F. & Doroshkevich, A.G., 1994, Ap.J., **420**, 26

Bardeen, J.M., Bond, J.R., Kaiser, N., & Szalay, A., 1986, Ap.J., **304**, 15 (BBKS)

Berezinsky, V. & Valle, J.W.F., 1993, Phys.Lett. **B 318**, 360

Bond, J.R., 1994, CMB workshop, Capri, Italy

Bond, J.R. & Efstathiou, G., 1984, Ap.J., **285**, L45

Bond, J.R. & Efstathiou, G., 1987, MNRAS, **226**, 655

Blumenthal, G.R., Faber, S.M., Primack, J.R. & Rees, M.J., 1984, Nature **311**, 517

Buryak, O.E., Demiański, M. & Doroshkevich, A.G., 1993, ApJ, **393**, 464

Buryak, O.E., Doroshkevich, A.G., Fong, R., 1994, Ap.J., **434**, 24

da Costa, L.N., Vogeley, M.S., Geller, M.J., Huchra, J.P. & Park, G., 1994, ApJ.Lett., **437**, L1

Demiański, M. & Doroshkevich, A.G., 1992, Int.J.Mod.Phys., **D1**, 303

Doroshkevich, A.G., 1984, Sov.Astron., **28**, 253

Doroshkevich, A.G., 1988, Sov.Astr.Lett., **14**, 125

Doroshkevich, A.G., 1989, Sov.Astron., **33**, 229

Doroshkevich, A.G., 1995, Proceedings of 11th Potsdam Workshop, in press

Doroshkevich, A.G., Fong, R., Gottloeber, S., Muecket, J.P. & Muller, V., 1995, MNRAS, submitted (DFGMM)

Doroshkevich, A.G., Fong, R., Gottloeber, S., Muecket, J.P. & Muller, V., 1995, in preparation

Doroshkevich, A.G. & Tucker, D., 1995, in preparation

Doroshkevich, A.G. & Turchaninov, V.I., 1995, Ap. J., submitted

Efstathiou, G., 1988, in Large-Scale Motions in the Universe. A Vatican Study Week, eds. V.C. Rubin & G.V. Coyne, Princeton University Press, Princeton, p. 299

Fong, R., Doroshkevich, A.G. & Turchaninov, V.I., 1995, in Bennett., C. et al., eds., Proc. 5th Maryland Astrophysics Conf., Dark Matter., AIP, New York, in press

Giavalisco, M., Macchetto, F.D., Madau, P. & Sparks, W.B., 1995, Ap. J. Let., **441**, L13

Gorski, K.M., Hinshaw, G., Banday, A.J., Bennett, C.L. et al., 1994, Ap. J. **430**, L89

Hu, W. & Sugiyama, N., 1994, Ap.J., submitted

Jorgensen, H.E., Kotok, E., Naselsky, P. & Novikov, I., 1995, Astron.& Astrophys, **294**, 639

Klypin, A.A., Holtsmann, J., Primak, J. & Regös, E., 1993 Ap. J., **416**, 1

Kofman, L., Pogosyan, D., Shandarin, S.F. & Melott, A.L. 1992, ApJ, **393**, 437

Partridge, R.B., 1988, Rep.Prog.Phys., **51**, 647

Peacock, J.A. & Dodds, S.J., 1994, MNRAS, **267**, 1020

Pebbles, P.J.E. & Yu, J.T., 1970, Ap.J., **162**, 815

Readhead, A.C.S., et al., 1992, Ann.Rev.Astron.Astrophys., **30**, 653

Shandarin, S.F. & Zeldovich, Ya.B. 1989, RMP, **61**, 185

Smoot, G.F., et al., 1992, ApJ. Lett., **396**, L1

Uson, J.M. & Wilkinson, D.T., 1984, Ap.J.Lett., **277**, L1, Nature, **312**, 427, Ap.J., **283**, 47

Vittorio, N. & Silk, J., 1984, Ap.J., **285** L39

White, M., Scott, D. & Silk, J., 1994, Ann.Rev.Astron.Astrophys., **32**, 319

Wilson, M.L. & Silk, J., 1981, Ap.J., **243**, 14

Zeldovich, Ya.B., 1970, Astron.& Astrophys., **5**, 20

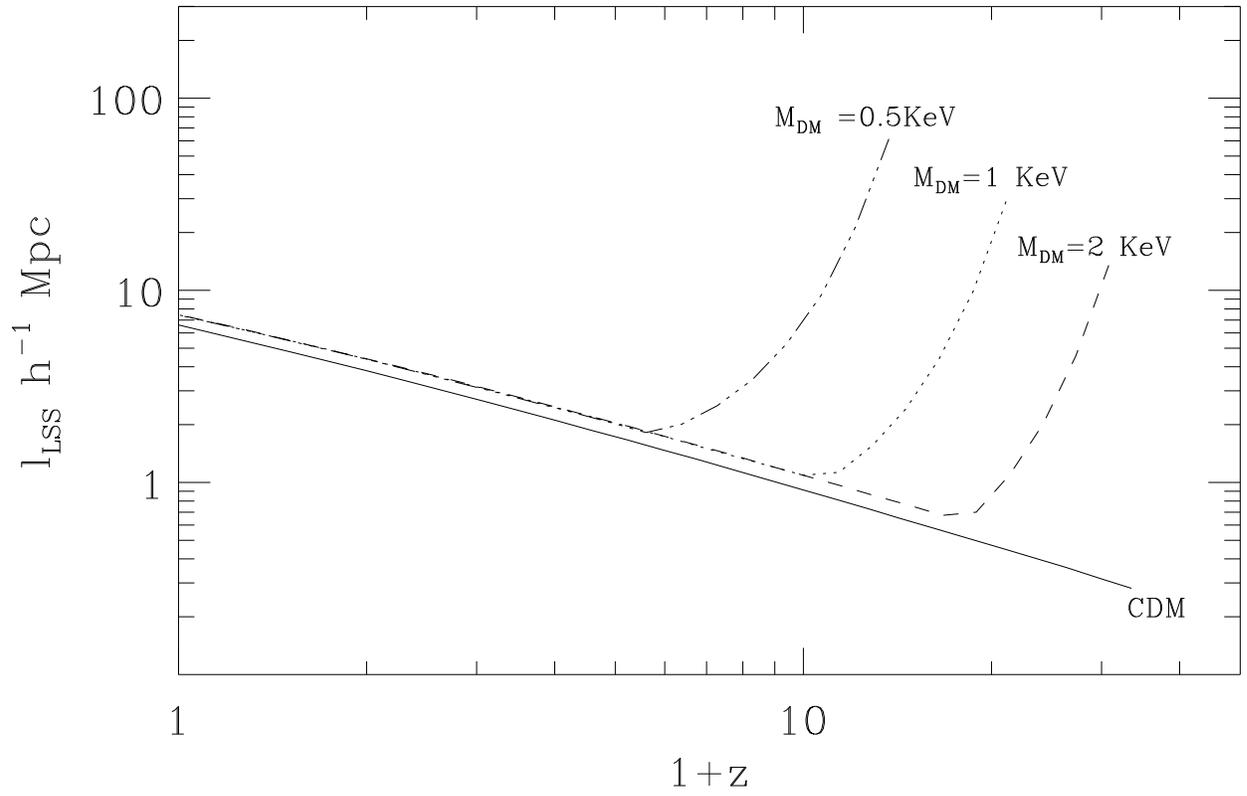

FIGURE 1. The mean separation of structure elements $l_{LSS}$ vs. redshift z for the scale−free CDM model and for WDM[1] models with various $M_{DM}$.

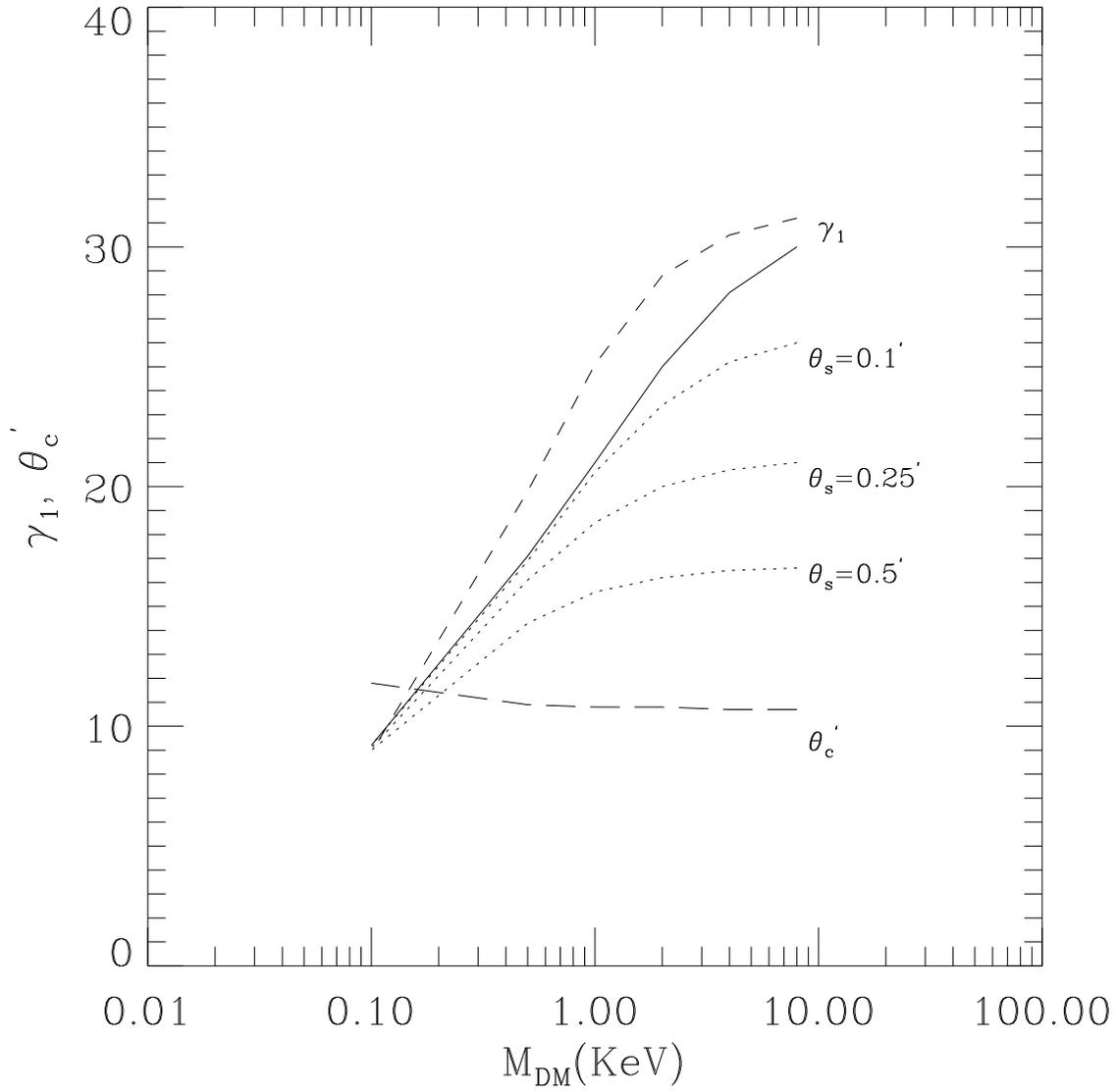

FIGURE 2. The coherence angle $\theta_c$ (arc min) and the factor $\gamma_1$ vs. mass of DM particle for WDM[1] (dashed line) and WDM[2] (solid line) for h=0.5 and $\Omega$=0.1. Dot lines show the antenna damping for different antenna beam−sizes, $\theta_s$.

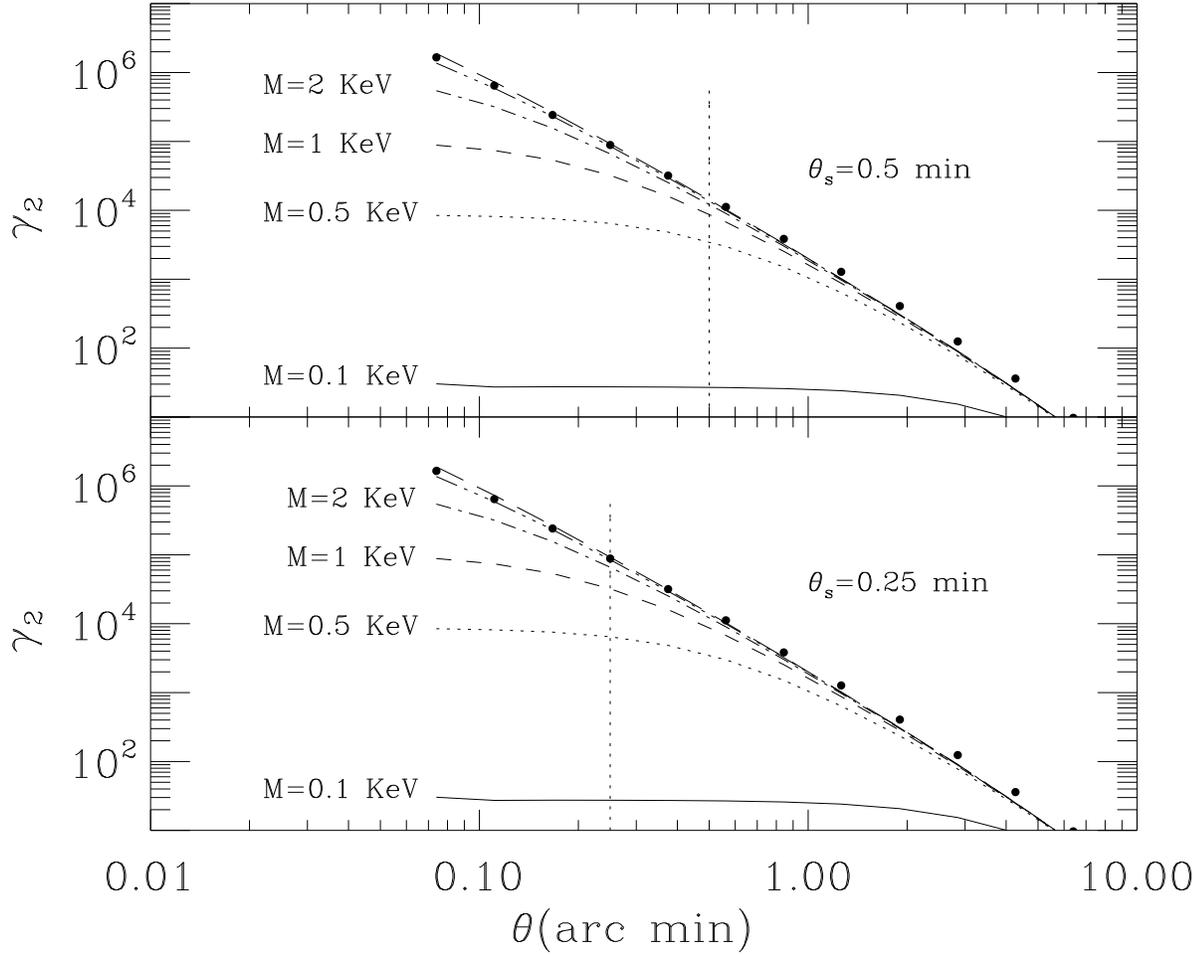

FIGURE 3a. The factor $\gamma_2$ vs. the switch angle $\theta$ for the scale-free CDM model (dotted) and for WDM[1] models with different masses of DM particles for h=0.5 and $\Omega_b$=0.1. Vertical dot lines show the antenna beam-size. The dashed-dot-dot and long-dashed lines correspond to $M_{DM}$=4 KeV and the $M_{DM}$=8 KeV.

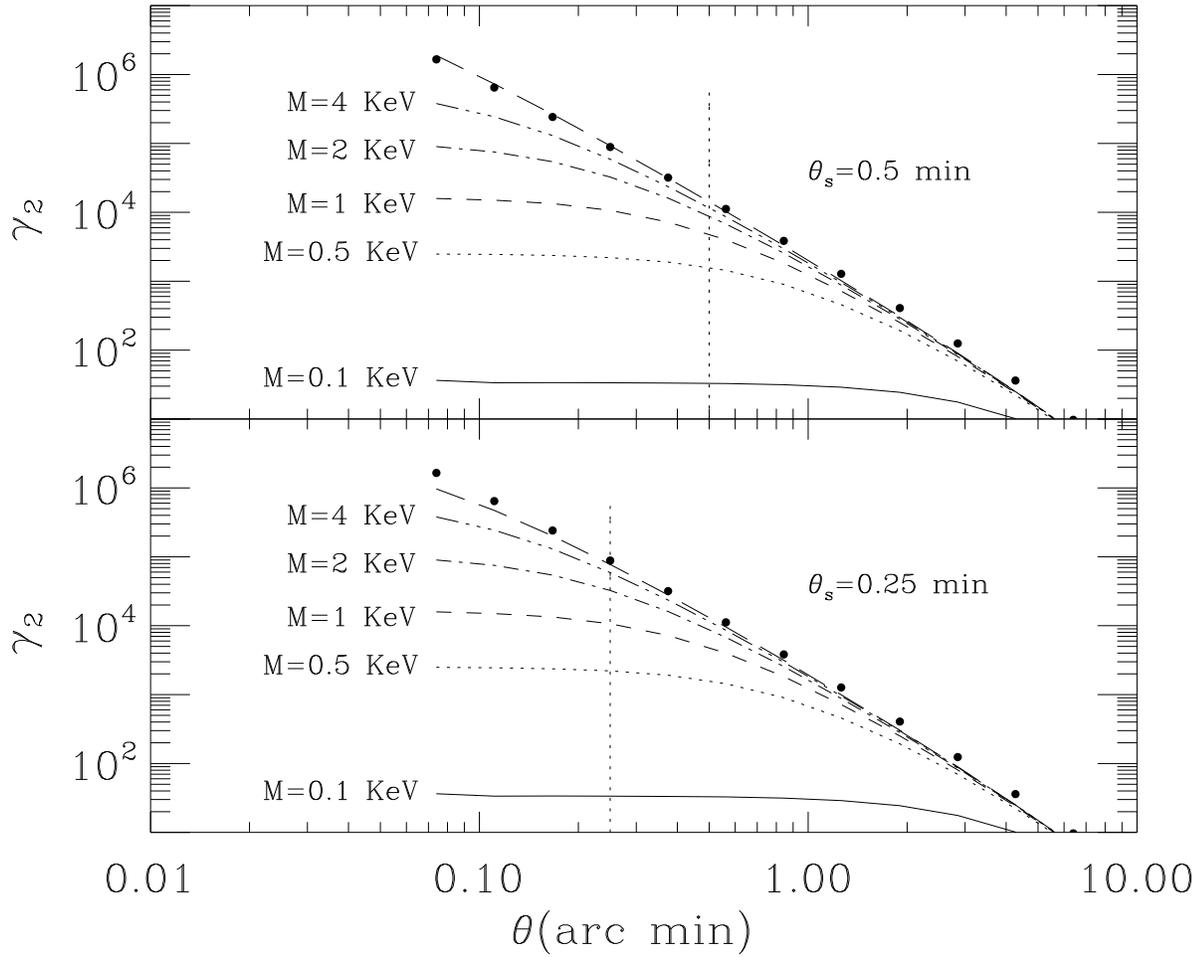

FIGURE 3b. The factor $\gamma_2$ vs. the switch angle $\theta$ for the scale-free CDM model (dotted) and for WDM$^{(2)}$ models with different masses of DM particles for $h=0.5$ and $\Omega_b=0.1$. Vertical dot lines show the antenna beam-size. The long-dashed line corresponds to $M_{DM}=8$ KeV.